\documentclass[hyper]{JHEP3}
\usepackage{amsmath}
\usepackage{amssymb}
\usepackage[utf8]{inputenc}
\usepackage{graphicx}
\usepackage{rotating}
\usepackage{setspace}
\usepackage{subeqnarray}
\usepackage{multirow}
\numberwithin{equation}{section}
\author{\"{O}mer Can G\"{u}rdo\u{g}an\\ \.{I}stanbul Teknik \"{U}niversitesi Fizik B\"{o}\"{u}m\"{u}, 34469, Maslak \.{I}stanbul, Turkey \thanks{Visiting Department of Physics, Swansea University, Swansea, SA2 8PP, UK} \email{gurdogano@itu.edu.tr}}
\abstract{A new solution in the string background dual to  $\mathcal{N}=1$ SQCD-like theories is presented. The gauge coupling in this solution has walking property. The Wilson loop calculations show that quark anti-quark potential makes phase transitions. Additionally the effect of flavours on other solutions in this background is investigated by considering some unflavoured solutions and perturbing them with small parameter $x=\frac{N_{f}}{N_{c}}$. }
\title{Walking solutions in the string background dual to $\mathcal{N}=1$ SQCD-like theories}
\begin{document}
%\maketitle
\tableofcontents
\section{Introduction}
The AdS/CFT correspondence \cite{Maldacena:1997re} conjectures an equivalence between string theory backgrounds and gauge field theories. The original duality proposes a correspondence between the  $3+1$ dimensional $\mathcal{N}=4$ supersymmetric Yang -- Mills (SYM) theory and type IIB string theory on $AdS_{5}\times S^{5}$. The conjecture is extended by further dualities between different string backgrounds and field theories with different properties than $\mathcal{N}=4$ SYM, such as spacetime dimensionality, less number of supersymmetries, etc. The string background studied in this work is a generalisation of the one proposed in \cite{Maldacena:2000yy} and utilised in \cite{Casero:2007jj} for the construction of a string dual to $\mathcal{N}=1$ supersymmetric QCD-like theories. 

The string duals of minimally supersymmetric QCD-like field theories employs D5-branes with world-volumes extended along the Minkowski directions and wrapped on a compact two -- sphere. Matter is added by the flavour D5-branes separated from the colour branes along a radial direction \cite{Casero:2006pt}. Then the theory includes matter fields in the fundamental representation, which is more suitable to phenomenological studies. Even though the exact solutions to this background are so far unknown, solutions for specific values of $N_{f}$ and $N_{c}$, possible asymptotics and numerical solutions are already available and their properties have been studied to some extent. For some papers where the flavoured MN background has been studied see \cite{Gaillard:2008wt,Caceres:2007mu, Casero:2007pz, Cotrone:2007qa, Bertoldi:2007sf}. For other papers dealing with flavours added by backreacting flavour branes see \cite{Benini:2006hh,Benini:2007gx, Ramallo:2008ew, Bigazzi:2008ie,Arean:2008az, Burrington:2007qd, Bigazzi:2008qq, Apreda:2006bu, Bigazzi:2005md,  Bigazzi:2008zt, Bigazzi:2009gu}.

One of the known solutions has the property that the coupling constant in its dual field theory has the so called walking feature \cite{Nunez:2008wi}. This is a required feature for the coupling between the techniquarks of technicolor models that are used as a natural model for the electroweak symmetry breaking. Similar to the case of QCD, the coupling constant diverges at the low energies. In this work a new numerical solution to the background with a walking gauge coupling is presented. The major difference to the one in \cite{Nunez:2008wi} is that the coupling constant remains bounded everywhere. This is achieved by adjusting some parameters so that the background functions have a certain asymptotical behaviour. These asymptotics have already been classified in \cite{Casero:2006pt,Casero:2007jj,HoyosBadajoz:2008fw}.

The second section is a review of the background with its known solutions. The properties of the known solutions is presented as well as the classification of the asymptotics. In the third section, the bounded walking solutions are constructed by imposing the solutions to have certain classes of asymptotics. It turns out to be that there is a constraint on the parameters of the system to produce the desired kind of solutions. In the fourth section, the Wilson loops have been calculated for the presented solution. The heavy quark potential exhibits a phase transition. Similar phase transitions of the heavy quark potential have been related to the phase transitions of the van der Waals gases \cite{Bigazzi:2008gd} and \cite{Bigazzi:2008qq}. Finally in the fifth section, not relevant to the bounded walking solutions, the effect of flavours is studied by finding flavoured solutions in a perturbative way.
\section{Review of the string dual to $\mathcal{N}=1$ SQCD-like theories}
\subsection{Description of the background}
In this chapter the background that is proposed to be dual to SQCD like theories will be reviewed very briefly. It is built on the Maldacena -- Nu\~{n}ez background, which is a type IIB supergravity solution introduced and proposed to be dual to a $\mathcal{N}=1$  SYM-like theory in \cite{Maldacena:2000yy} ( see also \cite{Chamseddine:1997nm} where a 4d solution was found). This theory includes $N_{c}$ colour D5-branes extending along the Minkowski directions $\vec{x}_{1,3}$ and the compact directions ($\theta$, $\phi$) that provide the colour symmetry of the theory. Then, $N_{f}$ flavour D5-branes smeared along four compact coordinates ($\theta$, $\varphi$, $\tilde{\theta}$, $\tilde{\varphi}$) are added to introduce an open string sector. Their worldvolume extends along the Minkowski directions, the radial coordinate that will be denoted as $\rho$ and a compact direction $\psi$. The open string sector gives rise to fundamental fields in the field theory. This  system is governed by the action $S=S_{IIB}+S_{flavour}$ which consists of the Type IIB supergravity action and the Dirac -- Born -- Infeld + Wess -- Zumino (DBI+WZ) action for the flavour branes. They read:
\begin{equation}
\begin{split}
S_{IIB}=&\frac{1}{2\kappa^{2}_{(10)}}\int d^{10}x\sqrt{-g}\left[R-\frac{1}{2}(\partial_{\mu}\phi)(\partial^{\mu}\phi) -\frac{1}{12}e^{\phi}F^{2}_{(3)}\right]\\
S_{flavour}=&T_{5}\displaystyle\sum^{N_{f}}\left(-\int_{\mathcal{M}_{6}}d^{6}xe^{\frac{\phi}{2}}\sqrt{-\hat{g}_{(6)}}+\int_{\mathcal{M}_{6}}P[C_{6}] \right)
\end{split}
\end{equation}
where $\hat{g}_{6}$ is the pullback of the metric and $P[C_{6}]$ s the pullback on the worldvolume of the RR six-form of the background. Having added the flavour branes, their effect on the background cannot be neglected, thus they are said to be backreacting. The action of the D5 flavour branes is six dimensional and they live on constant values of the two spheres' coordinates $(\theta, \phi)$ and $(\tilde{\theta}, \tilde{\phi})$. The smearing mentioned in the above paragraph avoids the need of delta functions for these six dimensional objects in the ten dimensional theory and this technique was first used in \cite{Bigazzi:2005md}. The ansatz for the metric of the string background is a generalisation of the background in \cite{Maldacena:2000yy}.  The metric describes a space of the topology $\mathbb{R}^{1,3}\times\mathbb{R}\times S^{2}\times S^{3}$  and it reads in the notation of \cite{HoyosBadajoz:2008fw}:
\begin{equation}
\begin{split}
%ds^{2}=\alpha'g_{s}N_{c}e^{\frac{\phi}{2}}\left[\frac{1}{\alpha'g_{s}N_{c}}dx_{1,3}^{2}+d\rho^{2}+e^{2h(\rho)}\left(d\theta^{2}+\sin^{2}(\theta)d\varphi^{2}\right)+\frac{e^{2g(\rho)}}{4}\left(\left(\tilde{\omega}_{1}  +a(\rho)d\theta\right)^{2}  \right. \right. \\ \left.\left. +\left(\tilde{\omega}_{2}+a(\rho)\sin(\theta)d\varphi\right)^{2}\right)+\frac{e^{2k(\rho)}}{4}\left(\tilde{\omega}_{3}+\cos(\theta)d\varphi\right)^{2}\right]
ds^{2}=&e^{\frac{\phi}{2}}\left[ dx_{1,3}^{2}+Y(\rho)\left(4 d\rho^{2} +(\omega_{3}+\tilde{\omega}_{3})^{2}\right)+\frac{1}{2}P(\rho)\sinh\left(\tau(\rho)\right)(\omega_{1}\tilde{\omega}_{1}-\omega_{2}\tilde{\omega}_{2})\right.\\
&+\left.\frac{1}{4}\left(P(\rho)\cosh\left(\tau(\rho)\right)+Q(\rho)\right)(\omega_{1}^{2}+\omega_{2}^{2})+\frac{1}{4} \left(P(\rho)\cosh\left(\tau(\rho)\right)-Q(\rho)\right)(\tilde{\omega}_{1}^{2}+\tilde{\omega}_{2}^{2})\right]
\end{split}
\end{equation}
The dilaton $\phi(\rho)$ is a function of the radial coordinate and the RR 3-form is given by
\begin{equation}
\begin{split}
	F_{(3)}=&\frac{N_{c}}{4} \left[ - \left(\tilde{\omega}_{1}+b(\rho)d\theta\right)\wedge\left(\tilde{\omega}_{2}-b(\rho)\sin{(\theta)d\varphi} \right)\wedge\left(\tilde{\omega}_{3} + \cos{(\theta)}d\varphi\right) \right.\\
	&\left.+b'(\rho)d\rho\wedge\left(-d\theta\wedge \tilde{\omega}_{1}+\sin{(\theta)}d\varphi\wedge\tilde{\omega}_{2}\right)+\left(1-b^{2}(\rho) \right)\sin{(\theta)d\theta}\wedge d\varphi\wedge\tilde{\omega}_{3} \right]\\
	&-\frac{N_{f}}{4}\sin{(\theta)}d\theta\wedge d\varphi\wedge\left(d\psi+\cos{(\tilde{\theta})d\tilde{\varphi}}\right)
\end{split}
\end{equation}
where
\begin{align}
	\begin{split}
		\tilde{\omega}_{1}&=\cos{(\psi)}d\tilde{\theta}+\sin{(\psi)}\sin{(\tilde{\theta})}d\tilde{\varphi},\\
		\tilde{\omega}_{2}&=-\sin{(\psi)}d\tilde{\theta}+\cos{(\psi)}\sin{(\tilde{\theta})}d\tilde{\varphi},\\
		\tilde{\omega}_{1}&=d\psi+\cos{(\tilde{\theta})}d\tilde{\varphi}.
	\end{split}
\end{align}
The background functions $P$, $Q$, $Y$, $\phi$, $a$ and $b$ are determined by a set of equations. These are BPS equations which are first order differential equations obtained from the supersymmetry transformations of the fermions and they ensure that the background satisfies the equations of motion of the gravity plus flavour branes action. After making some redefinitions like $a=\frac{P\sinh(\tau)}{P\cosh(\tau)-Q}$, $b=\frac{\sigma}{N_{c}}$ and the introduction of the function
\begin{equation}
\omega:=\sigma-\tanh(\tau)\left(Q+\frac{2N_{c-N_{f}}}{2}\right)
\end{equation}
the BPS equations are written as follows \cite{HoyosBadajoz:2008fw}:
\begin{align}
\label{BPS-P}
	\begin{split}
		P'& =8Y-N_{f},\\
		\partial_{\rho}\left(\frac{Q}{\cosh(\tau)}\right)&=\frac{(2N_{c}-N_{f})}{\cosh^{2}(\tau)}-\frac{2\omega}{P^{2}}(P^{2}-Q^{2})\tanh(\tau)\\
		\partial_{\rho}\left(\frac{\Phi}{\sqrt{P^2-Q^2}}\right)&=2\cosh(\tau)\\
		\partial_{\rho}\left(\frac{\Phi}{\sqrt{Y}}\right)&=\frac{16Y}{P^{2}-Q^{2}}\\
		\tau'(\rho)+2\sinh(\tau)&=-\frac{2Q\cosh(\tau)}{P^{2}}\omega\\
		\omega'&=\frac{2\omega}{P^{2}\cosh(\tau)}\left(P^{2}\sinh^{2}(\tau)+Q\left(Q+\frac{2N_{c}-N_{f}}{2}\right)\right)\\
	\end{split}
\end{align}
with an algebraic constraint $\omega=0$ and $\Phi\equiv \left(P^{2}-Q^{2}\right)Y^{\frac{1}{2}}e^{2\phi} $. One then calculates $\tau$ and $Q$ using (\ref{BPS-P}) as:
\begin{equation}
\begin{split}
\label{Q}
\cosh (\tau)&=\coth (2\rho)\\
Q&=\left(Q_{0}+\frac{2N_{c}-N_{f}}{2}\right)\cosh(\tau)+\frac{2N_{c}-N_{f}}{2}(2\rho\cosh(\tau)-1)
\end{split}
\end{equation}
These equations can be cast into a single differential equation for $P$ that reads \cite{HoyosBadajoz:2008fw}:
\begin{equation}
P''+(P'+N_{f})\left(\frac{P'+Q'+2N_{f}}{P-Q}+\frac{P'-Q'+N_{f}}{P+Q}-4\cosh(\tau)\right)=0.
\label{masterP}
\end{equation}
Once having obtained $P$, the other functions can be calculated in terms of it using the equations (\ref{BPS-P}). The equation (\ref{masterP}) will be  referred as the master equation. Its detailed derivation can be found in \cite{Nunez:2008wi, HoyosBadajoz:2008fw}.

\subsection{Type A backgrounds}
In particular, the backgrounds described above represent a relatively general case and they are called type N backgrounds in the notation of \cite{Casero:2007jj}. There is a special case in which the backgrounds are called type A. In this case, the fibration of the $S^{2}$ and the $S^{3}$ becomes simpler. On the field theory side, as opposed to type N backgrounds, the gaugino condensate in the field theory vanishes. It is possible to reproduce the type A backgrounds by setting $\sigma$=$\tau$=0. This unifying approach was presented in \cite{HoyosBadajoz:2008fw}. 

When dealing with type A backgrounds, it is useful to redefine the background functions:
\begin{equation}
\label{PQHGtrans}
\begin{split}
H=\frac{P+Q}{4}\\
G=\frac{P-Q}{4}
\end{split}
\end{equation}
In this notation the metric and the 3-form of the type A backgrounds are given by:
\begin{equation}
	\begin{split}
		ds^{2}=e^{\phi(\rho)}\left[ dx^{2}_{1,3}+4Y(\rho)d\rho^{2}+H(\rho)(d\theta^{2}+sin^{2}(\theta) d\varphi^{2})+G(\rho)(d\tilde{\theta}^{2}+sin^{2}\tilde{\theta}d\tilde{\varphi}^{2})\right.\\
		\left.+Y(\rho)(d\psi+\cos(\tilde{\theta})d\tilde{\phi}+\cos(\theta) d\varphi)^{2}\right]
	\end{split}
\end{equation}
\begin{equation}
	F_{(3)}=-d\left[ \sigma(\rho)(\omega_{1}\wedge\tilde{\omega}_{1}-\omega_{2}\wedge\tilde{\omega}_{2} \right]-\left(\frac{N_{f}-N_{c}}{4}\omega_{1}\wedge\omega_{2}+\frac{N_{c}}{4}\tilde{\omega}_{1}\wedge\tilde{\omega}_{2}\right)\wedge(\omega_{3}\wedge\tilde{\omega}_{3})
\end{equation}
with $\omega_{1}=d\theta$, $\omega_{2}=\sin(\theta)d\varphi$, $\omega_{3}=\cos(\theta)d\varphi$. The master equation in type A backgrounds reduces to
\begin{equation}
\label{Heq}
H''-\left(\frac{1}{2}\partial_{\rho}H+\frac{1}{4}(N_{f}-N_{c})\right)\left[-2\frac{\partial_{\rho}H+N_{f}-N_{c}}{H}-\frac{N_{f}+2\partial_{\rho H}}{H+\frac{N_{f}-2N_{c}}{2}\rho-C} \right]=0.
\end{equation}
As before, the other type A background functions can be calculated in terms of  $H(\rho)$ and integration constants \cite{Casero:2007jj}.
\begin{subeqnarray}
\label{BPS-H}
G=H+\frac{N_{f}-2N_{c}}{2}\rho-C\\
Y=\frac{1}{2}\partial_{\rho}H+\frac{1}{4}(N_{f}-N_{c})\\
\phi=\phi_{0}+\int\left[\frac{N_{f}-N_{c}}{4H}+\frac{N_{c}}{4G}\right]d\rho
\end{subeqnarray}

\subsection{Some known solutions}
Although the master equation is a highly non-linear second order differential equation, it permits simplifications by setting its parameters to certain values. For these specific cases there are some known analytic solutions. The ones that are considered in this work are listed below. 

While only $P(\rho)$ or $H(\rho)$ is given here, it is possible to calculate the remaining functions by the equations and the definitions of the previous section. For type N backgrounds see the equations (\ref{BPS-P} - \ref{Q}). Equations (\ref{BPS-H}) are valid for the type A case. It is always possible to translate between the ``$P$-$Q$'' and ``$H$-$G$'' notations using equations (\ref{PQHGtrans}).
\subsubsection*{Unflavoured solutions $N_{f}=0$}
\begin{itemize}
\item A type N solution from \cite{Maldacena:2000yy} is:
\begin{equation}
	P=2N_{c}(\rho-\rho_{0}),\ Q_{0}=-N_{c}-2N_{c}\rho_{0},\ \rho \geq \rho_{0} > -\infty
	\label{turo}
\end{equation}
\item  The  type A limit ($\tau, \sigma \rightarrow 0$) of the type N solution in eq. (\ref{turo}) above is:
	\begin{equation}
		H=N_{c}(\rho-\rho_{*}),\ \rho_{*}:=-\frac{1}{2N_{c}}\left(Q_{0}\frac{N_{c}}{2}\right),\ \rho \geq \rho_{*}
	\end{equation}
\end{itemize}
\subsubsection*{Solutions for $N_{f}=2N_{c}$}
$N_{c}=2N_{f}$ is a special case of the background where the functions have some different properties. In this case $Q(\rho)$ does not diverge unlike the other cases but it goes to a constant. 
\begin{itemize}
\item A type A solution called the ``conformal'' solution in \cite{Casero:2007jj} reads
\begin{equation}
H=\frac{N_{c}}{\xi}
\end{equation}
where $\xi$ is a real number between $0$ and $4$.
\item A deformation of the above solution
\begin{equation}
H=\frac{N_{c}}{16}(9\pm 3)+\frac{c_{+}}{4}e^{\frac{4\rho}{4}}
\end{equation}
also solves the master equation and it is of type A.
\end{itemize}
Further solutions for this case have been presented in \cite{Caceres:2007mu}.
\subsubsection*{An arbitrarily flavoured type A solution}
The following type A solution is an obvious solution of equation (\ref{Heq}). However it is not physical since the background function $Y(\rho)$ turns out to be zero (see eq. (\ref{BPS-H})).
\begin{equation}
\label{halfro}
H(\rho)=\frac{N_{f}-N_{c}}{2}(c_{1}-\rho)
\end{equation}
\subsection{IR and UV asymptotics and their classification}
Besides the exact solutions that are valid for specific cases, there are asymptotic expansions written for the generic cases. There are two kinds of asympotics for the UV and three for the IR limit. The UV asymptotics are referred as ``Class I and II'' while the IR ones are ``Type I, II and III''.
\subsubsection*{UV asymptotics}
As discussed in \cite{HoyosBadajoz:2008fw}, equation (3.17) of \cite{HoyosBadajoz:2008fw} implies that the $P$ must diverge as $\rho \rightarrow \infty$ unless $N_{f}=2N_{c}$. There are two possibilities. $P$ either asymptotes to $\propto \rho$ or $\propto e^{\frac{4\rho}{3}}$. They are called Class I and II UV behaviours, respectively.
\begin{itemize}
\item \bf Class I \rm In cases where $N_{f}\neq N_{c}$, the class I asymptotics are of two kinds depending on whether $N_{f}$ is larger or smaller than $2N_{c}$. For $N_{f}>2N_{c}$, $H(\rho)$ asymptotes to $\frac{1}{4}(N_{f}-N_{c})$, while for the other case it asymptotes to $\frac{1}{2}(2N_{c}-N_{f})\rho$. This class of asymptotics contains terms up to the first order in $\rho$. The leading term of $G$ and $Y$ as $\rho \rightarrow \infty$ is a constant, namely $\frac{N_{c}}{4}$. The dilation like $e^{4\phi}\propto e^{4\rho}/\rho$.  
\item \bf Class II \rm In class II solutions, $H(\rho)$ and $G(\rho)$ asymptote to $c_{+}e^{\frac{4\rho}{3}}$ regardless of the parameters. $G$, $Y$ are also proportional to $e^{\frac{4\rho}{3}}$ and the exponential of the dilaton 
$e^{4\phi}$ approaches a constant.
\end{itemize}
Since $\tau$ vanishes at the large values of $\rho$ , there is no distinction between type N and type A backgrounds in the UV region.
\subsubsection*{IR asymptotics}
The three types of IR behaviours are classified by the vanishing of some of the background functions in the IR region. The definition range of $\rho$, thus its smallest value called the IR value varies from case to case. The asymptotics of special type A backgrounds are reviewed here, since it is the relevant case for this work. In type II backgrounds the IR limit is either $\rho=0$ or $\rho=-\infty$.
\begin{itemize}
\item \bf Type I: \rm In such backgrounds the IR limit is defined as $\rho \rightarrow -\infty$. The series expansion for $H$ has the following form:
\begin{align}
H(\rho)=\frac{N_{f}-N_{c}}{2}(c_{1}-\rho)+\displaystyle\sum_{k\geq 1} \mathcal{P}_{k}(\rho)e^{4k\rho}
\end{align}
where $\mathcal{P}_{k}$ are polynomials of order $k+1$ in $\rho$. From this $Y$ is found to be proportional to $e^{\frac{4\rho}{3}}$. As stated above, the solution given in eq. (\ref{halfro}) is a bad solution because of the vanishing of $Y(\rho)\ \forall\rho$ . The type I solutions are solutions that do not suffer from this, however they approach eq. (\ref{halfro}) as $\rho\rightarrow \infty$. 
\item \bf Type II: \rm In backgrounds with type II IR behaviour, the IR limit is defined as $\rho \rightarrow 0$. The series expansion for the background functions $H$, $G$, $Y$, $\phi$ is either of the forms
\begin{align}
\begin{split}
H&=h_{1}\rho^{\frac{1}{2}}+\left(\frac{h_{1}^{2}}{3C}-\frac{N_{f}}{2}\right)\rho+\dotsb\\
G&=-C+h_{1}\rho^{\frac{1}{2}}+\left(\frac{h_{1}^{2}}{3C}-\frac{N_{f}}{2}\right)\rho+\dotsb\\
Y&=\frac{h_{1}}{4\rho^{\frac{1}{2}}}+\frac{1}{12}\left(\frac{2h_{1}^{2}}{C}+3N_{c}-3N_{f}\right)+\frac{3}{4}h_{1}\frac{72C^{2}+20h_{1}^{2}+3C(10N_{c}-7N_{f})}{72C^{2}}\rho^{\frac{1}{2}}+\dotsb\\
\phi&=\phi_{0}+\frac{N_{f}-N_{c}}{2h_{1}}\rho^{\frac{1}{2}}+\frac{3C(N_{f}-N_{c})^{2}-h_{1}^{2}(2N_{c}+N_{f})}{12Ch_{1}^{2}}\rho+...\ \ \ \ \ \rm{for}\ C<0\\
\rm{or}\\
H&=C+h_{1}\rho^{\frac{1}{2}}-\left(\frac{h_{1}^{2}}{3C}+\frac{N_{f}}{2}\right)\rho+\dotsb\\ 
G&=h_{1}\rho^{\frac{1}{2}}-\left(\frac{h_{1}^{2}}{3C}+\frac{N_{f}}{2}\right)\rho+\dotsb\ \\
Y&=\frac{h_{1}}{4\rho^{\frac{1}{2}}}-\frac{1}{2}\left(\frac{h_{1}^{2}}{C}+\frac{N_{c}}{2}\right)+h_{1}\frac{72C^{2}+20h_{1}^{2}+3C(10N_{c}-3N_{f})}{96C^{2}}\rho^{\frac{1}{2}}+\dotsb\\
\phi&=\phi_{0}+\frac{N_{c}}{2h_{1}}\rho^{\frac{1}{2}}+\frac{3C-h_{1}^{2}(3N_{f}-2N_{c})}{12Ch_{1}^{2}}\rho+\dotsb\ \ \ \ \ \rm{for}\ C<0\\
\end{split}
\end{align}
depending on the sign of C. Note that one of the functions vanishes at $\rho=0$ while the other is non-zero, $C$. For $C=0$, both of the functions are zero and these kind of backgrounds are called type III. Their leading behaviour is like $\rho^{\frac{1}{3}}$ as described below.
%\begin{eqnarray}
%\item \bf Type I \rm 
%H(\rho)=C+h_{1}\sqrt{\rho}-\left(\frac{h_{1}^{2}}{3C}\right)\rho+h_{1}\frac{72C^{2}+20h_{1}^{2}+30CN_{c}}{72C^{2}}\rho^{\frac{3}{2}}+...\\
%G(\rho)=h_{1}\sqrt{\rho}-\left(\frac{h_{1}^{2}}{3C}+N_{c}\right)\rho+h_{1}\frac{72C^{2}+20h_{1}^{2}+30CN_{c}}{72C^{2}}\rho^{\frac{3}{2}}+...\\
%Y(\rho)=\frac{h_{1}}{4\sqrt{\rho}}-\frac{1}{2}\left(\frac{h_{1}^{2}}{3C}+\frac{N_{c}}{2}\right)+h_{1}\frac{72C^{2}+20h_{1}^{2}+30CN_{c}}{72C^{2}}\rho^{\frac{3}{2}}+...\\
%\phi = \phi_{0}+\frac{N_{c}}{2h_{1}}\rho^{\frac{1}{2}}+\frac{3C-2h_{1}^{2}N_{c}}{12Ch_{1}^{2}}\rho+...
%\end{eqnarray}
\item \bf Type III: \rm Finally, there is the third kind of IR asymptotics that read. 
\begin{align}
\begin{split}
H&=h_{1}\rho^{\frac{1}{3}}+\frac{1}{10}(5N_{c}-7N_{f})\rho+\dotsb\\
G&=h_{1}\rho^{\frac{1}{3}}-\frac{1}{10}(5N_{c}+2N_{f})\rho+\dotsb
\end{split}
\end{align}
\end{itemize}
Type III asymptotics are valid instead of type II when $C=0$. Both $H$ and $G$ are zero in the IR limit.

The above review summarises the known possible asymptotical IR and UV behaviours of the solutions for the type A BPS equations of the background being studied. A new numerical solution found with class II UV and type II IR behaviour as well as the conditions for its existence and some of its implications on the field theory side such as the gauge coupling will be presented in the following section.
\section{Bounded walking solutions of the background}
%As reviewed in the previous chapter, the sAs stated above, the solution given in eq. \ref{halfro} is a bad solution because the vanishing of $Y(\rho)$. The type I solutions are solutions that do not suffer from this, however they approach eq. (\ref{halfro}) as $\rho\rightarrow \infty$. As stated above, the solution given in eq. \ref{halfro} is a bad solution because the vanishing of $Y(\rho)$. The type I solutions are solutions that do not suffer from this, however they approach eq. (\ref{halfro}) as $\rho\rightarrow \infty$. As stated above, the solution given in eq. \ref{halfro} is a bad solution because the vanishing of $Y(\rho)$. The type I solutions are solutions that do not suffer from this, however they approach eq. (\ref{halfro}) as $\rho\rightarrow \infty$. series expansions for IR and UV asymptotics of the solutions have been classified. In this chapter, the 0solutions of interest have type II IR and class II UV behaviour. That is, the solutions considered are of the form
%\begin{eqnarray*}
%H\propto C + \sqrt{\rho} & \textrm{for small } \rho\\
%H\propto e^{\frac{4\rho}{3}} & \textrm{for large } \rho \textrm{.}
%\end{eqnarray*}
Walking is a desired property of the so-called techniquark coupling constants of technicolor models that are used as a natural model for the electroweak symmetry breaking. Because of phenomenological reasons their running coupling constant must differ from the QCD coupling constant. The techniquarks need to remain strongly coupled for some higher energy scale. In other words, the coupling constant is needed to ``walk''  at a certain value before it decays to zero in the far UV. This kind of walking has already been observed in the background studied in this paper \cite{Nunez:2008wi}. The coupling constant can be translated for type A backgrounds from the background functions as (see \cite{Casero:2007jj}):
\begin{equation}
g^{2} \propto \frac{1}{H+G}.
\end{equation}
While the coupling constant in \cite{Nunez:2008wi} diverges in the IR like the QCD coupling, it can be useful to find backgrounds with bounded field theory coupling constants. $g^{2}$ in backgrounds with type II IR and class II UV asymptotics is expected to be constant for a certain interval and drop quickly at a certain point in the UV. That is indeed what happens at least for some values of the parameters $h_{1}$ and $C$ in the IR expansion.

\begin{figure}
\centering
\begin{tabular}{ccc}
\includegraphics[width=1.60in]{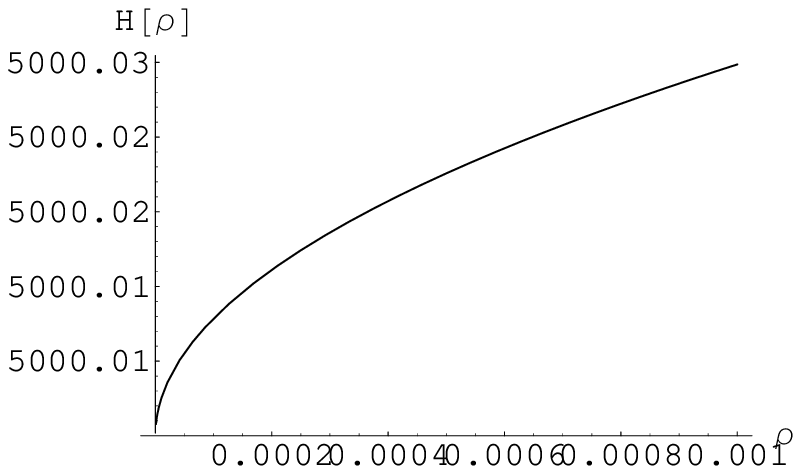}&\includegraphics[width=1.60in]{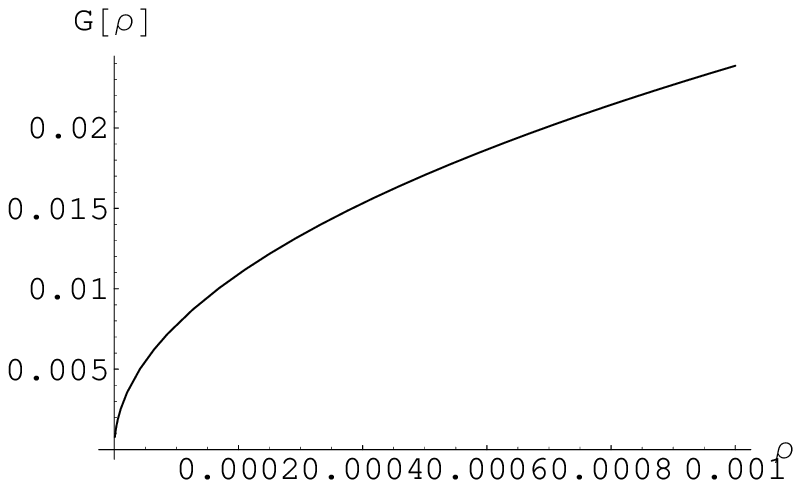}&\includegraphics[width=1.60in]{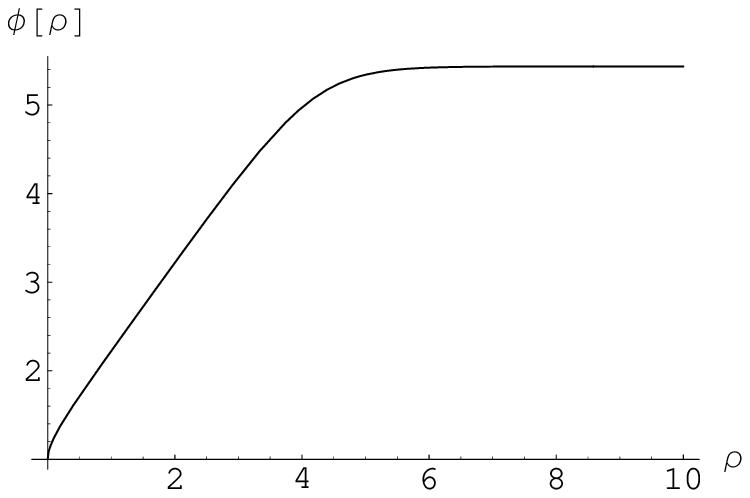}\\
\includegraphics[width=1.60in]{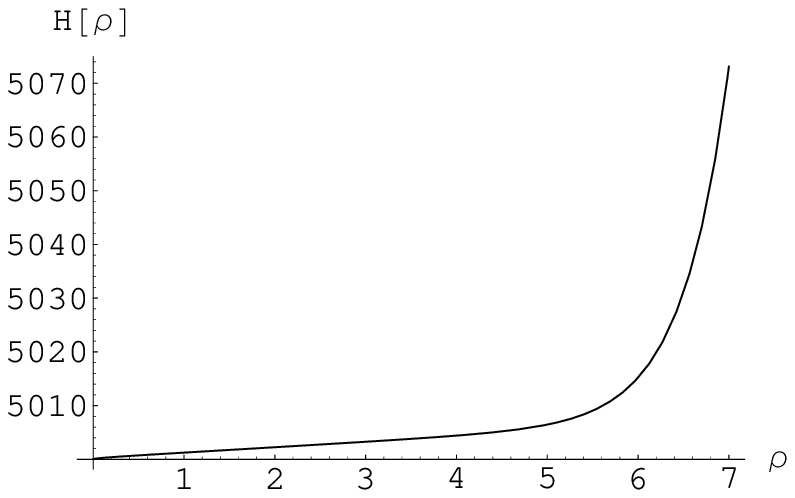}&\includegraphics[width=1.60in]{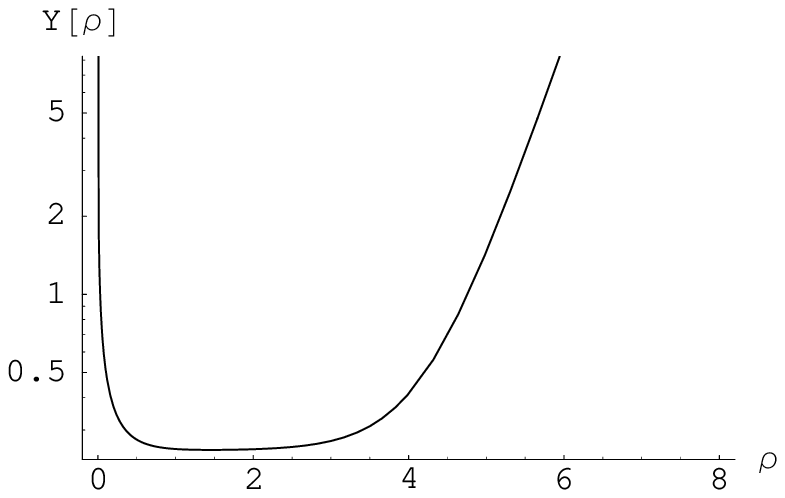}&\includegraphics[width=1.60in]{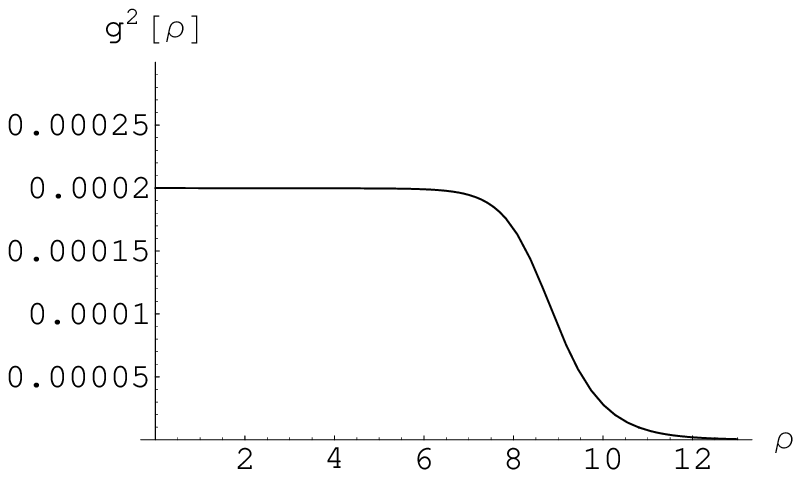}\\
\end{tabular}
\caption{A sample walking solution of the background with type II IR and Class II UV asymptotics. The plots are for $C=5000$, $h_{1}=0.7852$. There is no significant change in the coupling constant until the far UV values of $\rho$.}
\label{sampleplots}
\end{figure}

To investigate the parametrical constraints to obtain the desired kind of solutions a set of sample parameters $$(C,h_{1})\in \{1, 10, 50, 500, 5000\}\times\{0.01, 0.7, 0.78545, 0.78552, 0.78657, 0.85, 10, 15\}$$ has been considered where $N_{f}$ has been taken as zero at the first step. These solution is obtainable with some of the combinations of parameters from this set. Figure \ref{sampleplots} contains the plots of the background functions and the coupling constant for such a solution. The functions $H(\rho)$ and $G(\rho)$ have a square root behaviour in the IR while they diverge exponentially in the UV region. The dilation has no singularities and asymptotes to a constant in the UV. The coupling constant has the desired shape.
The table in figure \ref{cconstants} contains plots of the coupling constant for different parameters $C$ and $h_{1}$. As it can be seen from that table, the existence of the desired solutions depends on both of the integration constants. Moreover, the bound of one parameter is related to the value of the other.
\subsection{Parameter constraints for the desired kind of solutions}
\begin{figure}
\centering
%\begin{tabular}{r|ccc}
%&$g^{2}$&IR&UV\\
%\hline
%\hline
%C=1&\includegraphics[height=1.1in]{C1A.eps}&\includegraphics[height=1.1in]{C1B.eps}&\includegraphics[height=1.1in]{C1D.eps}\\
%\hline
%C=10&\includegraphics[height=1.1in]{C10A.eps}&\includegraphics[height=1.1in]{C10B.eps}&\includegraphics[height=1.1in]{C10D.eps}\\
%\hline
%C=50&\includegraphics[height=1.1in]{C50A.eps}&\includegraphics[height=1.1in]{C50B.eps}&\includegraphics[height=1.1in]{C50D.eps}\\
%\hline
%C=500&\includegraphics[height=1.1in]{C500A.eps}&\includegraphics[height=1.1in]{C500B.eps}&\includegraphics[height=1.1in]{C500D.eps}\\
%\hline
%C=5000&\includegraphics[height=1.1in]{C5000A.eps}&\includegraphics[height=1.1in]{C5000B.eps}&\includegraphics[height=1.1in]{C5000D.eps}\\
%\hline
%\end{tabular}\\
\begin{tabular}{cc}
\includegraphics[height=1.5in]{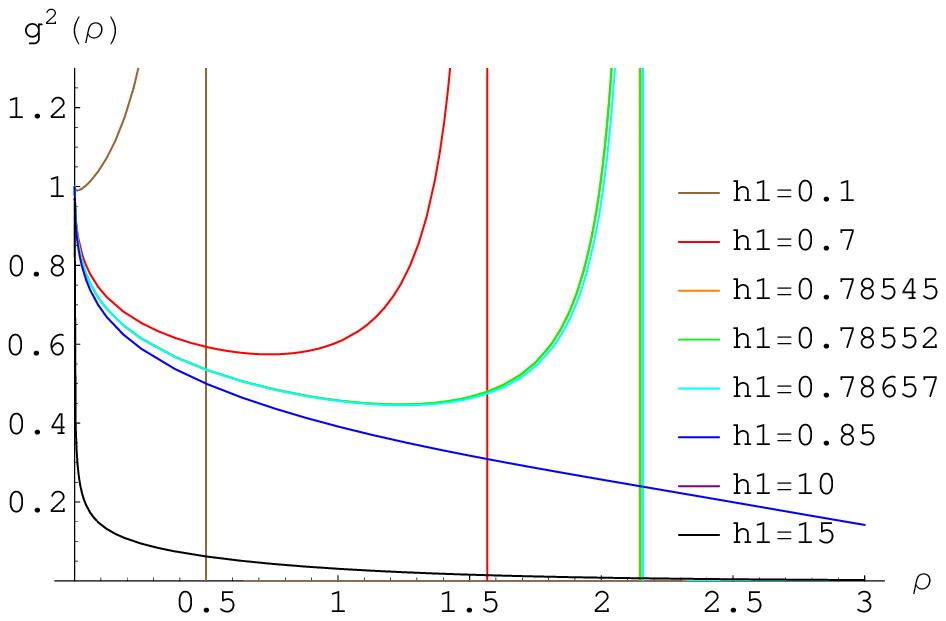}&\includegraphics[height=1.5in]{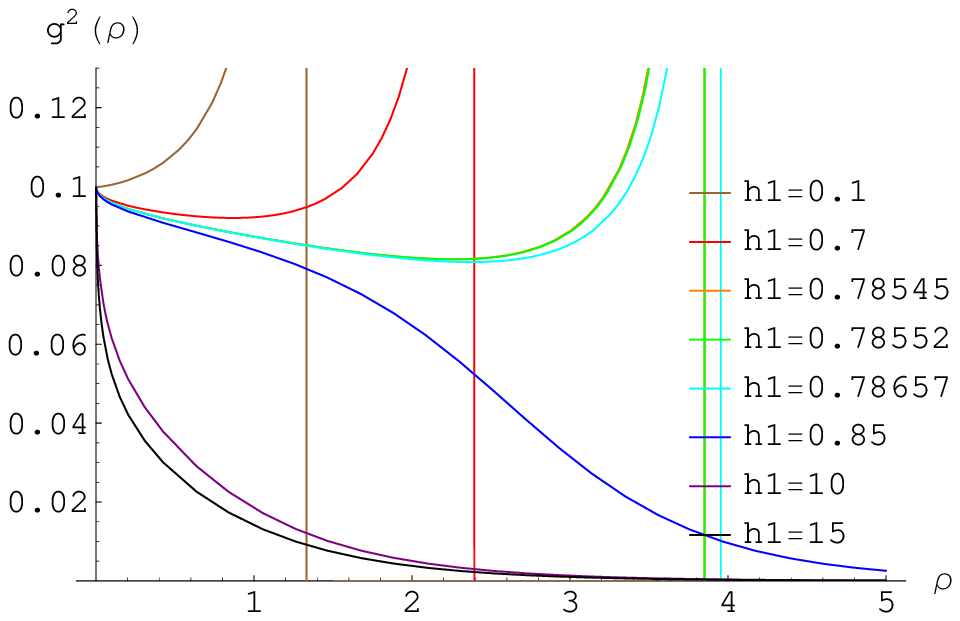}\\
C=1&C=10\\
\includegraphics[height=1.5in]{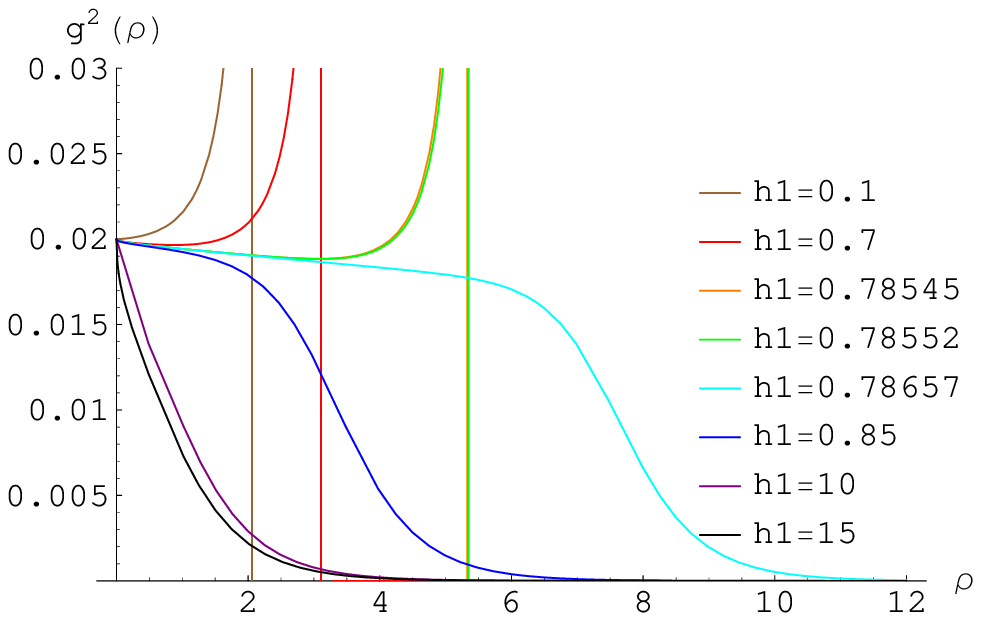}&\includegraphics[height=1.5in]{./C50.eps}\\
C=50&C=500\\
\includegraphics[height=1.5in]{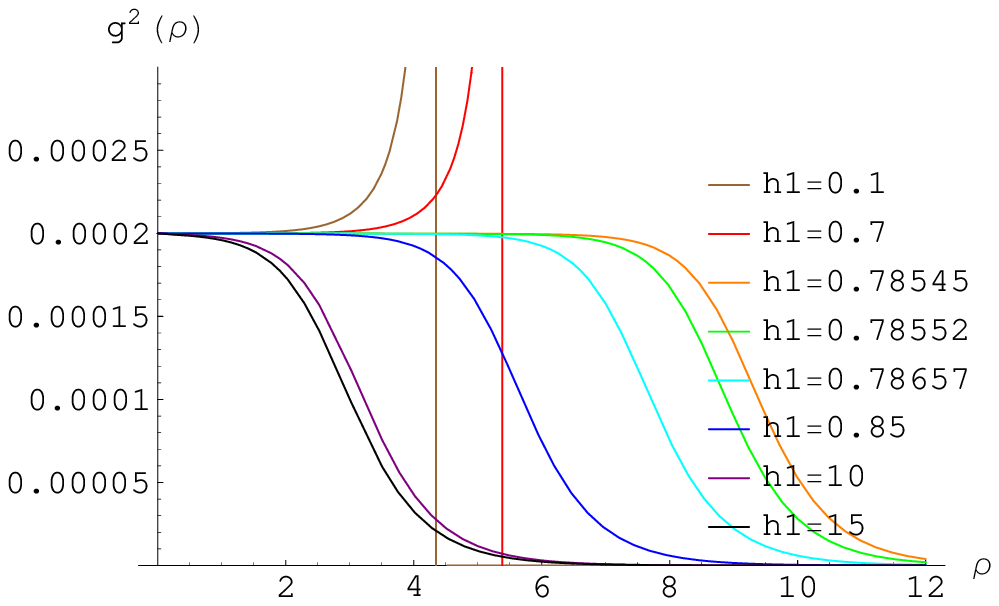}&\\
C=5000&\\
\end{tabular}
\caption{The plots of the field theory coupling constant $g^{2}$ for different values of $C$ and $h_{1}$ Each cell contains plots with common $C$. The colours represent the value of $h_{1}$
\label{cconstants}
}
\end{figure}

A more precise list of observations that can be concluded from figure \ref{cconstants} is the following:
\begin{itemize}
\item For solutions of common $h_{1}$, whose coupling constants blow at a certain point in the UV, become a smoothly and rapidly decaying behaviour as $C$ increases. At even larger $C$, they turn into Class II solutions ($\propto e^{\frac{4\rho}{3}}$). See the change of the yellow ($h_{1}=0.78552$) solution for different $C$.
\item For fixed $C$, smaller $h_{1}$ is required to have longer plateau. However, there is a minimum for $h_{1}$ that decreases as $C$ increases.
\item While for the ($h_{1}=0.85$) and ($h_{1}=0.78657$) lines $C=5000$ results in a longer plateau than $C=500$, for the solutions ($h_{1}=0.78552$) the case is opposite.
\end{itemize}
\subsubsection*{Adding $N_{f}$}
\begin{figure}
\centering
\begin{center}\includegraphics[height=1.8in]{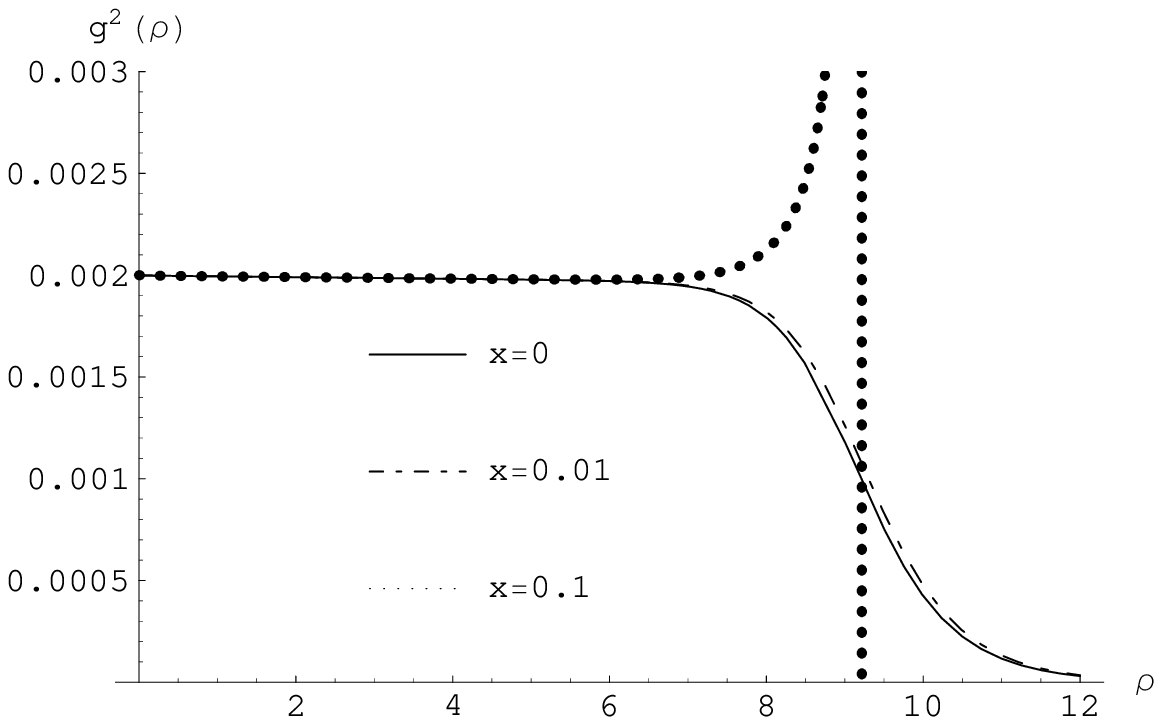}\includegraphics[height=1.8in]{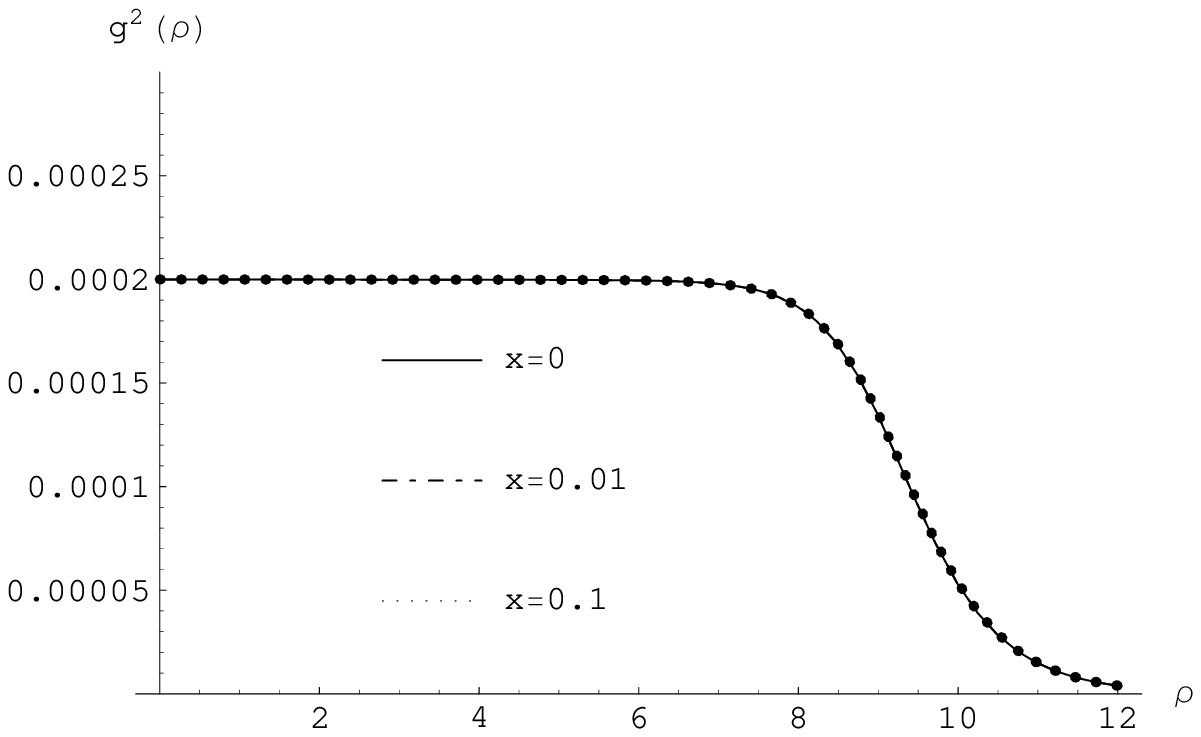}\end{center}
\caption{On the left hand side are the plots for $C=500$, the flavoured solutions do not look as desired. The right hand side plot  represents the solutions with $C=5000$ however, one sees that all of them are good and exactly the same.}
\label{fig:ccflavour}
\end{figure}

To get an intuition on the effect of flavours, a similar investigation has been made for flavoured solutions. Figure \ref{fig:ccflavour} contains plots of solutions with two different $C$ and constant $h_{1}=0.78659$. Different lines correspond to different $x\equiv\frac{N_{f}}{N_{c}} \in \{ 0,0.5,1\}$. 
Cases with $N_{f} > 0$ are only nice for larger $C$ than the cases with $N_{f}=0$. However, once $C$ is large enough the value of $x$ has no effect on the solutions.

\section{Wilson loops}
Wilson loops are gauge invariant operators that allow to compute the energy of an quark -- anti-quark pair. Their relation to gauge / gravity conjecture has been stated in \cite{Maldacena:1998im} and the calculation is reviewed by \cite{Sonnenschein:1999if} in detail. The Wilson Loop operator constructed over a closed space-time loop $\mathcal{C}$ representing the world-line of a quark -- anti-quark pair.  The world-line of pair we consider forms a rectangle where the pair is created in one side of it and annihilated in the other. When $T$, the time leg of the rectangle we consider is very large, the energy of the pair as a function of the separation of the pair can be estimated as follows:
\begin{equation}
\label{energy}
<W(\mathcal{C})>=A(L)e^{-TE(L)}
\end{equation}
On the other hand, it is proposed in \cite{Maldacena:1998im} that the expectation value of the Wilson loop operator can be calculated from the Nambu -- Goto action for a string stretching along the radial coordinate $\rho$. The endpoints of the string trace the quarks' world-lines living at the UV region (large $\rho$).
\begin{equation}
\label{wilsonev}
<W(\mathcal{C})>\propto e^{-S}
\end{equation}
\begin{equation}
S=\frac{1}{2\pi \alpha'}\int d\tau d\sigma \sqrt{|det(h_{\alpha\beta})|}.
\end{equation}
where $h_{\alpha \beta}=g_{\mu\nu}\partial_{\alpha}X^{\mu}\partial_{\beta}X^{\nu}$ is the induced metric on the string world-sheet.
After fixing the world sheet parametrisation to $\tau = t$ and $\sigma = x$), the determinant of the induced metric for the static string stretched to along the $\rho$ - $x$ plane is:
\begin{eqnarray}
det(h_{\alpha \beta})& =& g_{tt}g_{xx}+g_{tt}g_{\rho\rho} \left(\partial_{x}\rho\right)^{2}\\
		&=&\alpha '^{2} e^{2\phi}\left(-1-4Y\rho'^{2}\right) \nonumber
\end{eqnarray}
where $\rho '=\frac{d\rho}{dx}$.
Since there is no $t$ dependence in the integrant, the $t$ integration can be performed straight away to obtain:
\begin{equation}
\label{action}
S=\frac{T}{2\pi}\int dx e^{\phi}\sqrt{1+Y\rho '^{2}}=\frac{T}{2\pi}\int d \mathcal{L}
\end{equation}
The Lagrangian does not depend explicitly on $x$, therefore the $x$ -- Hamiltonian 
\begin{equation}
\mathcal{H}=\frac{\partial \mathcal{L}}{\partial \rho'}\rho'-\mathcal{L}
\end{equation}
is a constant of motion. In fact, it is equal to $-e^{\phi(\rho_{0})}$ where $\rho_{0}$ is the point on which $\rho'$ and consequently $\frac{\partial \mathcal{L}}{\partial{\rho'}}$ vanishes. Using this it is possible to write
\begin{equation}
\frac{d\rho}{dx}=\frac{1}{2\sqrt{Y(\rho)}}\frac{\sqrt{e^{2\phi(\rho)}+e^{2\phi(\rho_{0})}}}{e^{\phi(\rho_{0})}}
\end{equation}
to express the quark separation in terms of $\rho_{0}$, the deepest coordinate that the string reaches:
\begin{equation}
L=\int dx=2\int_{\rho_{0}}^{\rho_{1}} 2\sqrt{Y(\rho)}\frac{e^{\phi(\rho_{0})}}{\sqrt{e^{2\phi(\rho)}+e^{2\phi(\rho_{0})}}}d\rho.
\end{equation}
Similarly, writing the energy of the pair from the equations \eqref{energy}, \eqref{wilsonev} as a function of $\rho_{0}$ as
\begin{equation}
V=\int_{\rho{0}}^{\rho_{1}} \mathcal{L} \frac{dx}{d\rho} d\rho - 2\times\int_{0}^{\rho_{1}}2\sqrt{Y(\rho)}e^{\phi(\rho)}d\rho
\label{VQQ}
\end{equation}
one obtains a parametric relation between the quark -- anti-quark separation and energy. In equation (\ref{VQQ}), two quark masses calculated as the length of a string stretching from the bottom of the space to the quarks, have been subtracted from the energy to consider just the potential energy of the pair. 

Here we calculate the Wilson loops on a background with bounded walking coupling constants described previously. We take the loop to be the on the $x$ -- $t$ plane at a very large radial coordinate $\rho=\rho_{1}$.
\begin{figure}
\centering
\includegraphics[width=2.3in]{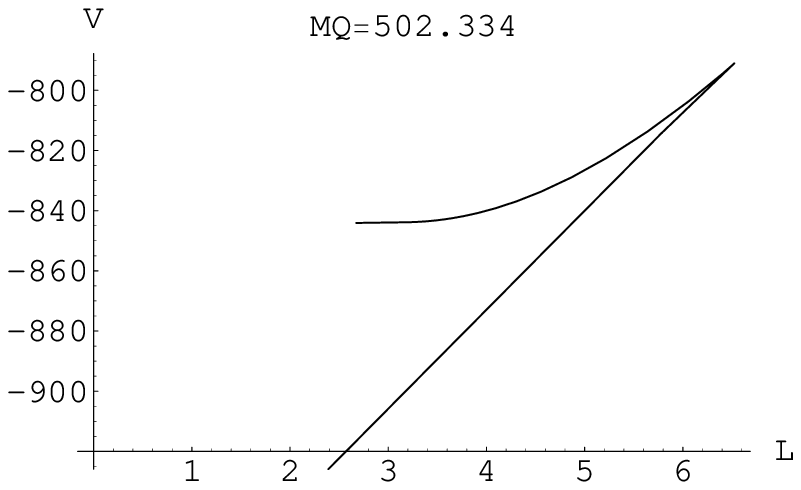}
\includegraphics[width=2.3in]{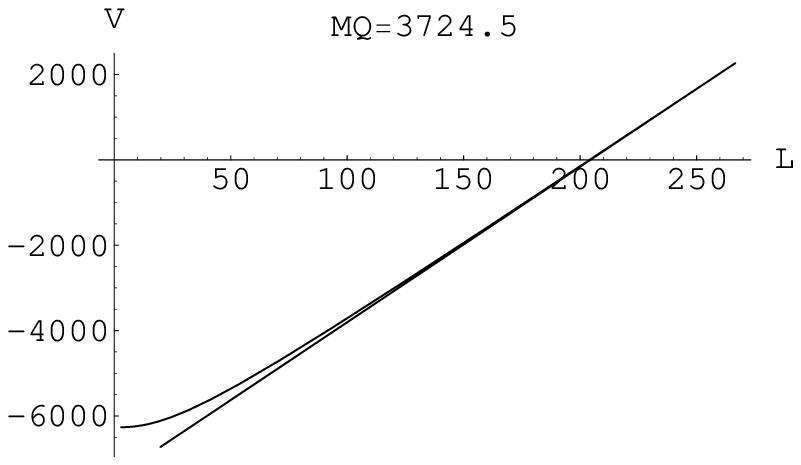}
\caption{The plot on the left shows the energy -- quark separation calculated using Wilson loop for quarks of masses $m_{Q}\approx 500$. The on the right hand side is the same for $m_{Q}=3700$.}
\label{fig:VL}
\end{figure}
Figure \ref{fig:VL} shows the plots of energy -- separation relation for configurations with quark masses $m_{Q}\approx500$ and $m_{Q}\approx3700$. One sees two branches and the lower one has linear confinement. The upper non-linear branch is less stable due to its larger energy. The branches coincide at a certain point at a very high $\rho_{0}\lesssim \rho_{1}$ after which the string cannot stretch any further. This is interpreted as string breaking. Even though there are no flavour branes in this background to allow formation of dynamical quarks where the string breaks, the derivative of the background function $H(\rho)$ is singular at the origin. That affects the energy and the separation integrals and thus is said to allow phenomena as if there were flavour branes.

Such behaviours have been observed and discussed in similar solutions of this background, see for example \cite{Casero:2007jj, Bigazzi:2008gd, Bigazzi:2008cc}. In \cite{Bigazzi:2008gd}, \cite{Bigazzi:2008cc} and \cite{Bigazzi:2008qq} flavoured systems with massive dynamical quarks are found out to have the energy - separation relation that qualitatively resembles the phase transition in the Gibbs free energy -- pressure relation of van der Waals gases that involves ``quasi-static phases''. Therefore, it has been argued \cite{Bigazzi:2008qq} that it can be interesting to study quark potentials in gauge theories with string duals by relating them to the phase transitions of van der Waals gases.

\begin{figure}
\centering
\label{wilsonsacma}
\includegraphics[height=1.5in]{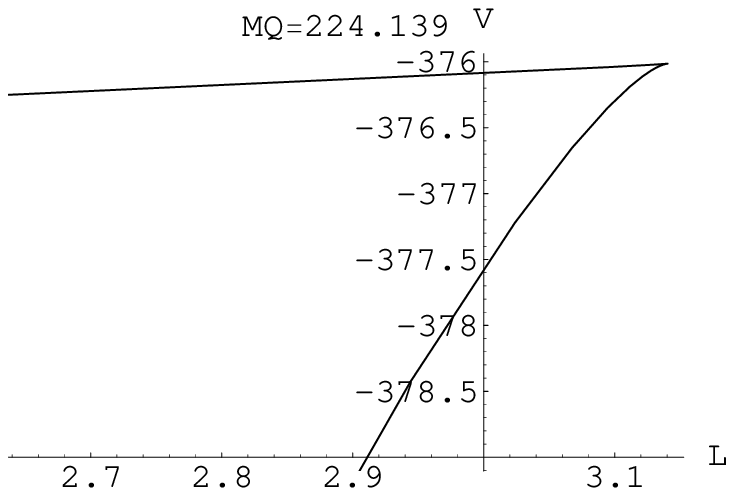}
\includegraphics[height=1.5in]{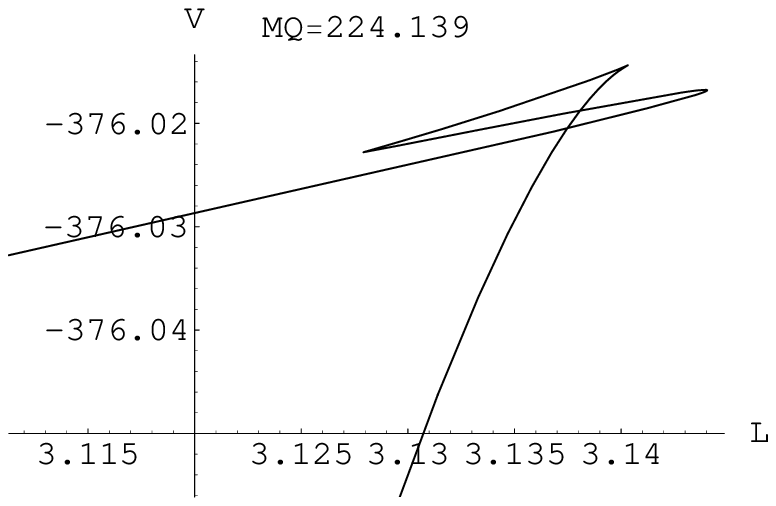}
\caption{Decreasing the heavy quark masses, the phase transition at $\rho \lesssim \rho_{1}$ is replaced by another transition at a lower value of $\rho$. The plot to the left shows the general picture while a zoom of the transition region is presented on the right hand side.}
\end{figure}

In the present background, another interesting phase transition occurs if one puts the brane where the heavy quarks live on at distances of smaller values of $\rho$. In other words, one reduces the mass of the quarks. Then a phase transition occurs at the configurations with about $\rho \approx 0.1\rho_{1}$. Figure \ref{wilsonsacma} contains the plot of the entire picture and the zoom of the region of transition

\begin{figure}
\centering
\label{wilsonVrho0}
\includegraphics[height=1in]{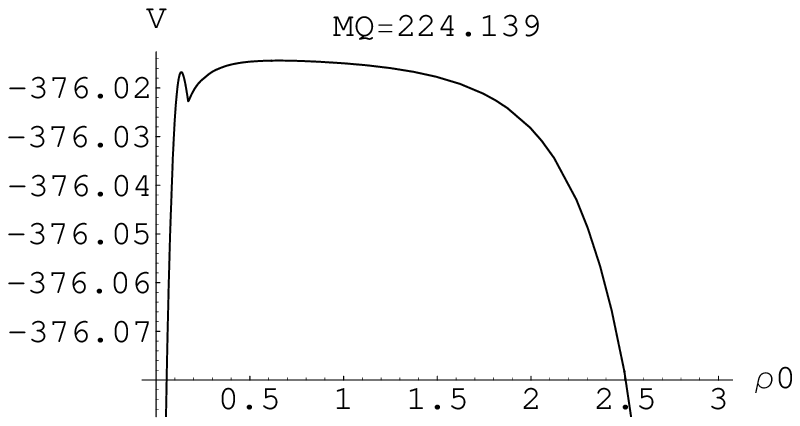}
\includegraphics[height=1in]{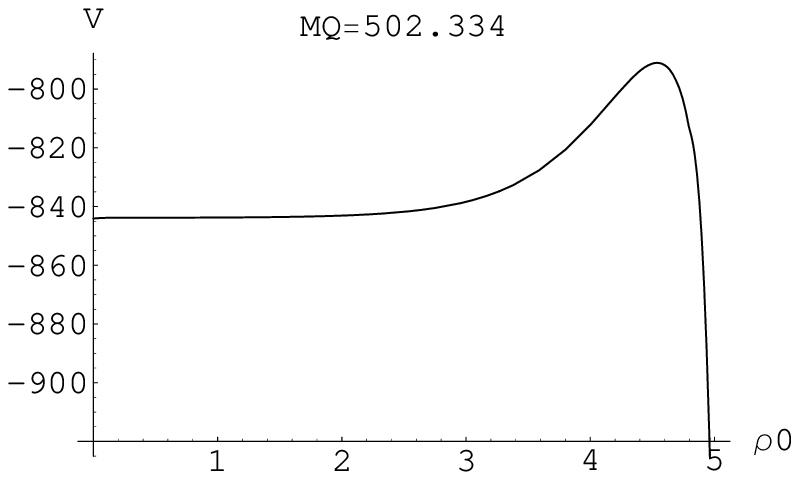}
\includegraphics[height=1in]{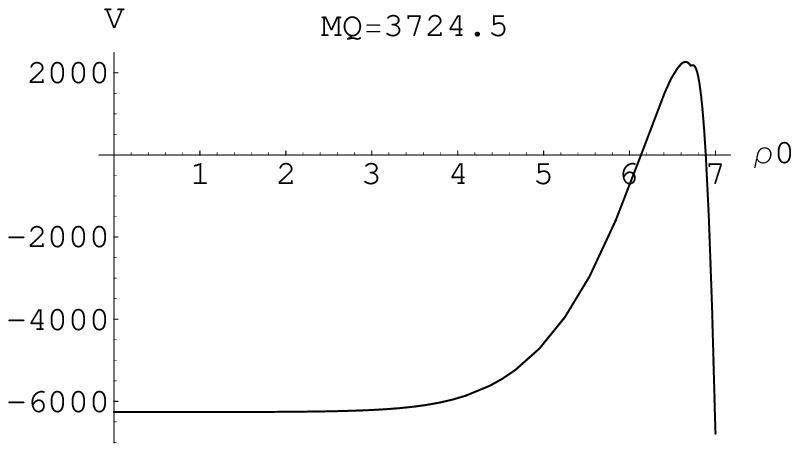}
\caption{ \small The relation between the quark potential energy $V$ and the deepest coordinate that the string reaches $\rho_{0}$. The plot to the left belongs to the case with $M_{Q}\approx 3700$, the plot in the centre belongs to the case with $M_{Q}\approx 500$ while the one to the right belongs to the case with $M_{Q}\approx 225$}|
\end{figure}

Figure \ref{wilsonVrho0} shows the relation between the quark potential and the deepest coordinate that the string reaches for the three cases mentioned previously in this chapter. The phase transitions occur at the points where the potential reaches a maximum or a minimum. 

Concluding this chapter, it is possible to say that one observes phase transitions of the heavy quark potential energy also in the bounded walking solutions of the background we presented in this work. Where the transition occur and what their qualitative properties are, depends on where one places the heavy quarks. None of the phase transitions here is exactly of the same type as the one of the van der Waals gases. However, there is still a possible relation since the pattern matches partially.
\section{Perturbative addition of flavours}
For the non-flavoured cases there are exact solutions known to the background functions. The idea of this section is to look for small $x=\frac{N_{f}}{N_{c}}$ expansions of some flavoured solutions, whose $x \rightarrow 0$ limit are the known non-flavoured solutions. In other words, we perturb the non-flavoured solutions around $x=0$ and plug the expression
\begin{equation}
H(\rho)=H_{0}(\rho)+h_{1}(\rho)x+h_{2}(\rho)x^{2}+...
\end{equation}
into the $H$ - equation and solve the differential equations obtained for $h_{n}(\rho)$ functions. The observation is that the properties of the solutions change with the addition of flavours.
\subsection*{Backgrounds with $H(\rho)=\frac{\rho}{2}$}
As mentioned around equation (\ref{halfro}), this solutions are known to be unphysical \cite{Casero:2007jj} since they have vanishing background function $Y=0$ for all $\rho$ and therefore cannot be considered as a 10 dimensional background. However, as soon as one adds flavour corrections up to the $k$th order in $x$, one obtains correction terms of the form 
$$\mathcal{P}_{k}(\rho)\times e^{4k\rho}.$$
where $\mathcal{P}_{k}$ are polynomials of order $k+1$. This solutions are as reviewed before known as type I and they are well-behaved solutions. Thus we say, a solution that is not even a background turns into a physical background after the addition of flavours. That the solution $H=\rho/2$ being the limit of type I solutions as $\rho \rightarrow -\infty$ has already been discussed in \cite{Casero:2007jj}.
\subsection*{Backgrounds with $H(\rho)=\rho$}
Another solution to the H-equation is $H(\rho)=\rho$. From this solution one calculates a dilaton that diverges at $\rho=0$. Diverging dilatons in the IR are considered as bad singularities causing unphysical effects - see section 5 of \cite{Maldacena:2000mw} . That is because $e^{\phi}$ has to be finite in the IR limit. Following the perturbative flavouring  procedure described above, one considers correction terms that come in powers of $x$ and solves the differential equations for the coefficients that are functions of $\rho$. Then one obtains the solutions of the form known as type II. In type II solutions the dilaton does not have a bad singularity  \cite{HoyosBadajoz:2008fw}. An example for a type II dilaton is depicted in figure \ref{sampleplots}.
\subsection*{Backgrounds with $P=2\rho$}
\begin{figure}
\centering
\includegraphics[height=1.5in]{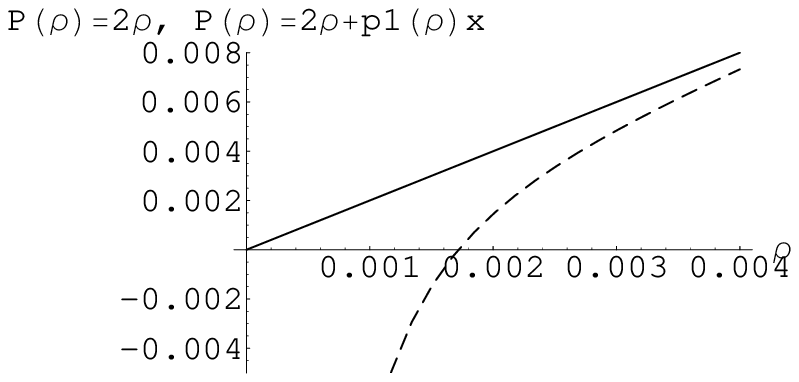}
\includegraphics[height=1.5in]{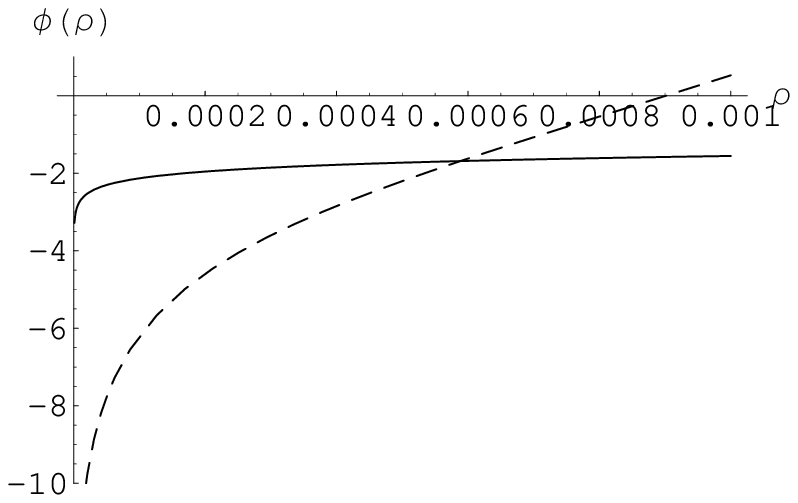}
\caption{The change of the background $P(\rho)=2\rho$ with flavours: The plot to the left shows $P$ and the one to the right shows $\phi$. The solid line is the unflavoured solution while the dashed line is the solution perturbed with solutions}
\label{hpq}
\end{figure}

This unflavoured type N solution is known to be a smooth solution that has no bad singularities. Perturbing this solution with flavours brings it to the following form (up to the first order in $x$ and second order in $\rho$).
\begin{equation}
P(\rho)=2\rho+x\left(\frac{c_{2}}{\rho^{2}}+c_{1}\rho-\frac{1}{2} +\mathcal{O}(\rho^{2})\right)+\mathcal{O}(x^{2})
\end{equation}
As usual, the dilaton is calculated from the equations (\ref{BPS-P}) as:
\begin{equation}
e^{4\phi}\propto\frac{1}{(P^{2}-Q^{2})Y\sinh^{2}(\tau)}
%\phi\propto\sqrt{P^{2}-Q^{2}}\sinh (2\rho)
\end{equation}
One sees that the dilaton diverges to $-\infty$ in the IR after adding the flavours. This divergence is not considered as bad since $e^{\phi}$ remains bounded. Figure \ref{hpq} contains the plots of the background functions $P$ and the dilaton $\phi$ before and after the perturbative addition of flavours.
\section{Conclusions}
In this work a solution to the string background has been presented which is dual to SQCD-like field theory with a gauge coupling that has a walking property. It has been observed that the integration constants that come from the BPS equations must be be constrained by a relation between each other in order to obtain such kind of backgrounds with walking gauge couplings. After obtaining the exact numerical solutions it is possible to make calculations for the quantities of the field theory such as the quark -- anti-quark potential using the Wilson loops. The Wilson loop calculations pointed out that the heavy quark potential for the field theory experiences phase transitions. The qualitative properties of those phase transitions depend on the quark masses which depends on the radial coordinate of the D-brane on which the quarks live. Possibly these phase transitions can be linked to the phase transitions of van der Waals gases by analogy. Finally, in addition to the calculations related to the presented solution, the effect of the presence flavours has been investigated. Some already known unflavoured exact solutions have been flavoured by a small $x=\frac{N_{f}}{N_{c}}$ perturbations. This procedure has cured the certain problems of the unflavoured backgrounds and generated previously known smooth solutions.
\section*{Acknowledgements}
It is a pleasure to thank the physics department of the University of Wales, Swansea where I was kindly hosted for the realisation of the present work. I am grateful to Dr. Carlos N\'{u}\~{n}ez for his invaluable guidance during this work. I also would like to thank Dr. Ioannis Papadimitriou and Dr. Maurizio Piai for the related discussions.
\bibliographystyle{JHEP-2}
\bibliography{0303}
\end{document}